\documentclass{ptptex}

\usepackage{graphicx}
\usepackage{wrapft}
\def\be{\begin{equation}}
\def\en{\end{equation}}

\def\gdot{\dot{\gamma}}

\def\gs{\gtrsim}
\def\ls{\lesssim}
\newcommand{\bi}[1]{\mbox{\boldmath$#1$}}
\newcommand{\bm}[1]{\mbox{\boldmath$#1$}}

\def\bea{\begin{eqnarray}}
\def\ena{\end{eqnarray}}

\title{
Jammed Particle Configurations and 
Dynamics in  High-Density 
Lennard-Jones Binary Mixtures  in Two Dimensions 
}

\author{Hayato Shiba$^1$ and Akira Onuki$^2$
}

\inst{$^1$Institute for Solid State Physics, University of Tokyo, Chiba 277-8581, Japan\\
$^2$Department of Physics, Kyoto University, Kyoto 606-8502,
Japan
}

\abst{
We examine  the changeover in the  particle 
configurations and the dynamics in dense Lennard-Jones binary mixtures composed of small and large particles.  By varying the composition 
 at a low temperature, we realize 
crystal with defects, polycrystal 
with small grains, and glass with various degrees of disorder. 
In particular, we show configurations 
where small crystalline regions composed of the majority species 
are enclosed by percolated amorphous layers composed 
of the two species. 
We visualize the dynamics of configuration changes 
 using the method of bond breakage 
and following the particle displacements.  
In  quiescent jammed states, the dynamics  is severely slowed down and is 
highly heterogeneous at any compositions. 
In shear,  plastic deformations 
multiply occur in relatively fragile regions, 
growing into large-scale   shear bands 
where the strain is highly localized. 
Such  bands appear on  short time scales 
and change on long time scales 
 with finite life times. }

\begin{document}

\maketitle

\section{Introduction}

Binary particle systems  with size dispersity 
exhibit complicated phase behavior 
depending  on the temperature $T$, the average number density 
$n=(N_1+N_2)/V$, and the composition 
$c=N_2/(N_1+N_2)$ \cite{Dick,Madden,Likos,Ito,Stanley,hama}. 
 Here $N_1$ and $N_2$ 
are the  numbers of the small and large particles  in a volume $V$, 
respectively.  At high densities, the particle configurations 
 sensitively depend on 
the size ratio $\sigma_2/\sigma_1$  between 
 the diameters of the two species, $\sigma_1$ and  $\sigma_2$.  
If $\sigma_2/\sigma_1$  is close to unity at  large $n$ 
and  at low $T$, a crystal state is realized 
with a small number of defects.  
However, if $\sigma_2/\sigma_1$  considerably deviates 
 from unity, crystal states are realized only for 
very small  $c$ or $1-c$. 
For not small $c$ and  $1-c$, polycrystal and glass states 
emerge without long-range crystalline order at low $T$.  
 It is then of great interest how the particle 
configurations and the dynamics 
change with varying the composition $c$ 
at considerably large size dispersity. 
As the glass transition is approached,  
the structural relaxation time $\tau_\alpha(T)$ 
grows  dramatically from a microscopic to macroscopic time,  
while  the particle configurations remain random 
yielding the structure factors similar 
to those in liquid.

The dynamics of   supercooled liquid and glass 
have been studied extensively 
using molecular dynamics simulations (MD). As a marked feature, 
the  glass dynamics  is  highly heterogeneous 
\cite{Takeuchi,Hiwatari,Harrowell,yo,Kob,Dol,Ha,Kawasaki,HamaOnuki,preprint2}.
In particular, Yamamoto and one of 
the present authors \cite{yo}  
examined breakage of appropriately 
defined bonds. 
The broken  bonds accumulated in long  time intervals 
 are analogous to the critical fluctuations in 
Ising systems  such that their 
structure factor may be fitted to 
the Ornstein-Zernike form,
\be  
S_{\rm b}(k)=S_{\rm b}(0)
/(1+k^2\xi^2),
\en 
in two dimensions (2D) and 
in three dimensions (3D). 
The  wave number $k$ 
is smaller than the inverse  
particle size.    The  correlation length $\xi$ can 
thus be  determined, which   grows  with 
lowering  $T$. Bond breakage events 
tend to take place repeatedly  in relatively fragile   regions,  leading to 
 aggregation of broken bonds on long time scales. 
 Kob  {\it et al.} \cite{Kob} 
pointed out relevance of  stringlike clusters of mobile 
particles whose lengths increase at low $T$. 
In addition, the diffusion constant in glassy 
materials is strongly affected by the dynamic heterogeneity 
\cite{Silescu,Ediger,yo2}. Afterwards, 
some authors have claimed the presence of 
  correlation  between 
 the  structural heterogeneity in the particle configurations 
and  the dynamic heterogeneity on long time scales 
\cite{Ha,HamaOnuki,Kawasaki}. Recently significant heterogeneity 
has  been found in the elastic moduli in glass 
\cite{Yoshimoto,Barrat-small}, which is the origin 
of  nonaffine  elastic displacements  
for very  small strains.

Glass dynamics  under shear flow 
has  also been studied by many authors 
\cite{yo1,An,Cates,Barrat-band,Ber,Barrat2D,Falk2D,Lama,Furukawa,preprint1}, 
where the shear rate $\gdot$ 
is a new  parameter representing the degree of nonequilibrium.   
Similar jamming rheology has been found in foam and  microemulsion systems, 
colloid suspensions, and granular materials 
\cite{Okuzono,Durian,Behringer,Da,hat}.  
In jammed states of molecular glass, each particle undergoes 
 shear-induced  configuration  changes  
 on the time scale of $\gdot^{-1}$. At a low temperature $T$, 
the  total  rate of the configuration change  is 
given by \cite{yo1,An}
\be  
\tau_\alpha (T,\gdot)^{-1}= 
\tau_\alpha(T)^{-1}+ A_\alpha\gdot.
\en   
In the right hand side, 
 the first term  is 
the thermal activation rate and the second one  is 
the shear-induced  rate, where the coefficient 
 $A_\alpha$ is    of order unity. 
The steady-state viscosity $\eta(T,\gdot)$ is  proportional to     
$\tau_\alpha (T,\gdot)$.  The following  dynamic scaling relation 
for the correlation length 
was in accord with the simulation  \cite{yo,yo1}:   
\be 
\xi(T,\gdot)  \sim 
\tau_\alpha (T,\gdot)^{1/z}, 
\en 
including the sheared case. The dynamic exponent 
$z$ was estimated as  $4$   in 2D and $2$ in 3D.  
In glass a Weissenberg number may be defined by  
\be  
{\rm Wi}= \tau_\alpha(T) \gdot.  
\en  
A nonlinear response  regime emerges  for  Wi$>1$. 
Around the glass transition,  $\tau_\alpha(T)$ 
grows and  the regime Wi$\gg 1$    
is realized   even for extremely small  shear,  where 
the configuration change is mostly induced by shear 
and $\eta \sim \gdot^{-1}$. 
Indeed in  this condition,  shear-thinning 
behavior was measured    in  glass 
under uniaxial stress \cite{Li,Simmons}. 
 Also  for   near-critical  and    
complex  fluids, the criterion of 
nonlinear shear  effects is  given by 
Wi$=\tau \gdot >1$ with an 
appropriate long relaxation time $\tau$ 
 \cite{Onukireview}.  Moreover, a number of MD simulations 
  of  sheared glass  have realized  self-organization of 
 ``shear bands" with high strain localization 
extending througout the system.  In simple 
shear flow such  bands are nearly 
 along the flow or the velocity-gradient 
direction   \cite{Barrat-band,Falk2D,Furukawa,preprint1}, while 
under uniaxial stress they make an angle of $\pi/4$ 
with respect to the uniaxial axis 
 \cite{Falk3D,bandB,Robbins,Bulatov,Onuki-plastic}. 
In metallurgy, shear bands have been observed in 
 amorphous solids under uniaxial stretching or compression 
 above a yield stress 
\cite{Acta,exp-band,Jing}.

Plastic deformations 
in crystal and polycrystal  are extremely complex and 
are still  poorly understood 
  despite their extensive 
research 
\cite{Friedel,dislocation,post,KubinD,Minami,Yip,Yama,Jack,HamanakaShear}. 
In crystal with low-density defects, 
dislocation motions play a major role in plasticity, where  they  
yield  slip planes with various sizes as observed in  
acoustic emission experiments \cite{dislocation} 
and by transmission electron microscopy \cite{post}. 
In polycrystal,  the particles in the vicinity of 
grain boundaries are relatively  mobile compared to those 
within the grains, so their collective motions give rise to 
 sliding  of the grain boundaries 
under applied stress \cite{Yip,Jack,HamanakaShear}. 
In these systems,  plastic events take place  as  
bursts or avalanches spanning  wide ranges of space and time 
scales. In this manner, they intermittently  release 
the elastic  energy stored in the crystalline regions 
at high strain.   On the other hand, in glass under  stress,  plastic 
events  have  been assumed  to be spatially 
localized due to the 
structural disorder in glass \cite{Spaepen,Argon,Langer}, 
but our recent 2D simulation 
has shown  that they frequently take place   over wide areas 
in short times  \cite{preprint1}.  
In glass, slip elements (slip lines in 2D)     
do not much exceed the particle  size, 
but they successively appear  in their neigborhood 
 forming  large-scale 
aggregates. When glass rheology is studied in  simulation,  
such  avalanches manifest themselves in   large 
  stress drops in the stress-strain curve. 
The  irregularity of the average stress versus time     
 has been conspicuous  in  
numerous  simulations in the literature, even though 
the curve should become 
smoother  with increasing the system size. 
Molecular 
glassy materials behave as  elastic bodies on short time scales 
 with mesoscopically  inhomogeneous elastic moduli 
   \cite{Yoshimoto,Barrat-small}. Thus 
large-scale release of the elastic energy  at high stress 
is a common   feature of  plasticity   
in crystal, polycystal, and glass.  

In this paper, we 
will investigate the heterogeneity  dynamics 
in high-density and low-temperature 
2D model binary mixtures
 without and with applied shear for various compositions 
on the basis of our previous work \cite{hama,HamaOnuki,HamanakaShear}.
We are interested in the relationship  between 
the dynamic heterogeneity on long time scales 
 and the  structural heterogeneity in the particle configurations. 
In a separate paper we have examined rheology and  
plastic deformations on  
various time scales \cite{preprint1}.

The organization of this paper is as follows.
In Sec.II,   our  model  
 and  our simulation method will be explained. 
 In Sec.III, numerical results 
will be presented on 
the particle configurations and the heterogeneous 
 dynamics for various $c$ without shear. 
 In Sec.IV, we will examine 
the collective dynamics under shear, which are 
even more heterogeneous than in quiescent states.

\section{ Background of Simulation} 

\subsection{Model}

Our two-dimensional (2D)  binary mixtures \cite{hama,HamaOnuki,HamanakaShear} 
consist   of two  species    1 and 2 interacting   via  
truncated Lennard-Jones (LJ) potentials,  
\begin{equation}
v_{\alpha\beta} (r) = 4\epsilon \left[ \left(\frac{\sigma_{\alpha\beta}}{r}\right)^{12} - \left(\frac{\sigma_{\alpha\beta}}{r}\right)^6\right] -C_{\alpha\beta} \quad
\label{eq:LJP}
\end{equation}
which are characterized by the energy $\epsilon$ 
and the interaction lengths    
$\sigma_{\alpha\beta} = 
(\sigma_\alpha +\sigma_\beta )/2$ $(\alpha,\beta=1,2)$. 
The $r=|{\bi r}_j-{\bi r}_k|$  is the particle 
distance. Hereafter ${\bi r}_j$  
denote the particle positions. 
The  diameters of the two species  are 
$\sigma_1$ and $\sigma_2$ and 
their ratio is fixed at  $\sigma_2/\sigma_1=1.4$. 
For $r>r_{\scriptsize{\textrm{cut}}} 
=3.2\sigma_1$, we set $v_{\alpha\beta} =0$
and the constant $C_{\alpha\beta}$ 
ensures the continuity of $v_{\alpha\beta}$
at $r=r_{\scriptsize{\textrm{cut}}}$. 
The particle number is   $N=N_1+N_2=$ {{9000}} 
and the  composition   
$c=N_2/N$ is varied. The system is in the region $-L/2<x,y<L/2$. 
The volume   $ L^2$ 
 is chosen such that the volume fraction 
of the soft-core regions is fixed as 
$(N_1\sigma_1^2 +N_2\sigma_2^2)/L^2 = 1$.
We integrated   the  equations of motion 
 using the leapfrog method, where  
the time step  is  $0.002\tau$ with
\begin{equation}
\tau = \sigma_1 \sqrt{m_1/\epsilon}.
\end{equation}
The  mass ratio is fixed at 
 $m_1/m_2 = (\sigma_1/\sigma_2)^2$. The following data were taken at 
$T=0.2\epsilon/k_B$. 
The space coordinates $x$ and $y$, the time $t$,  
the shear rate $\dot\gamma$,  
and the temperature $T$ will be measured 
in units of $\sigma_1$,  $\tau$, $\tau^{-1}$, 
 and $\epsilon /k_B$, respectively.

There was no tendency of phase separation in  our simulation. 
For the parameters we adopt in this paper, 
the structural relaxation time $\tau_\alpha(T,c)$ 
without shear is of order $10^4$ in glass, 
is longer  in polycrystal,   
and tends to  infinity in crystal  from the decay of 
the self-time-correlation function \cite{HamaOnuki}. 
The  grain boundary motions are  
severely slowed down 
in the presence of large  size dispersity, while 
they move rather fast in one-component systems. 
Thus $\tau_\alpha$ is longer  
 in polycrystal than in glass.

\subsection{Orientation angle  and disorder variable}

In the particle configurations at high density, 
a large fraction of the particles are 
enclosed by six particles.  
The local crystalline order may then  be 
represented by a sixfold orientation\cite{NelsonTEXT}. 
We define an orientation angle $\alpha_j$ in the range 
$-\pi/6 \le \alpha_j <\pi/6$ 
for each particle $j \in \alpha$ using the complex number 
 \cite{hama,HamaOnuki,HamanakaShear}, 
\be
\Psi_j =
\sum_{k\in\textrm{\scriptsize{bonded}}} \exp [6i\theta_{jk}]
=  |\Psi_j| e^{6i\alpha_j},
\label{eq:Alpha}
\en 
where the summation is over the ``bonded'' particle 
$k\in \beta$ satisfying 
$r_{jk} =|\bm{r}_j -\bm{r}_k|<1.5\sigma_{\alpha\beta}$.  
The $\theta_{jk}$ is the angle of the relative vector
$\bm{r}_j -\bm{r}_k$ with 
respect to the $x$ axis.   

We  also 
introduce  another non-negative-definite 
variable, called the disorder variable, 
which  represents the degree of the deviation from 
 the hexagonal order for each particles $j$ by
\be
D_j = \sum_{k\in\textrm{\scriptsize{bonded}}}
|e^{6i\alpha_j}-e^{6i\alpha_k}|^2= 
2 \sum_{k\in\textrm{\scriptsize{bonded}}} [1-\cos 6(\alpha_j -\alpha_k)].
\label{eq:Disodrv}
\en  
\begin{wrapfigure}{l}{6.6cm}
\centerline{\includegraphics[width=6.5cm,height=4.2 cm]{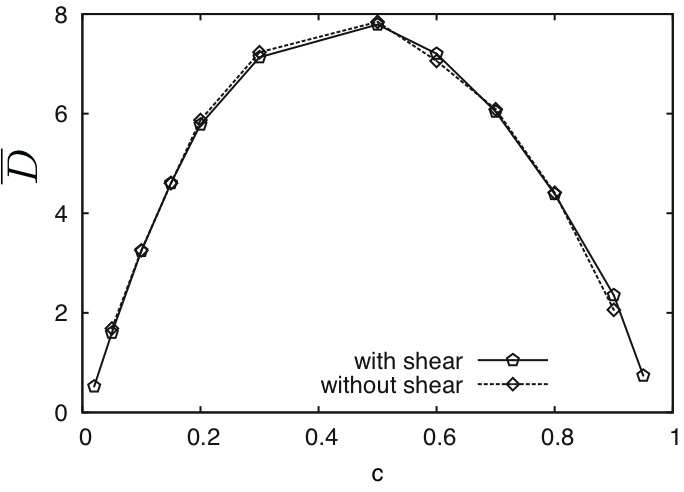}}
 \caption{Average disorder variables $\bar{D}$ vs $c$ 
without shear and with shear $\gdot=10^{-4}$, where $T=0.2$ 
and $N=9000$. 
 }
\end{wrapfigure} 
  Here $D_j$ are 
in the range 15-20 for particles around defects, 
while $D_j \cong 0$  for a perfect crystal at low $T$. 
The degree of the overall disorder  
may be represented by the average over the particles\cite{hama}, 
\be 
{\bar D}= \sum_{j=1}^N  D_j/N,
\en 
which is large in glass and liquid and is 
small in crystal. 
In Fig. 1, we show the average $\bar D$ 
versus $c$ at $T=0.2$ without shear and with shear 
$\gdot=10^{-4}$, where the two curves nearly coincide. 
For  this  shear rate 
(much smaller than the
 molecular frequency $\tau^{-1}$), there is no 
proliferation of defects induced by shear, 
while the defect structure is continuously deformed by shear.  
The maximum of $\bar D$ is attained at $c\sim 0.5$. 
See Figs. 8 and 9 of Ref. 6 for $\bar D$ 
as a function of $T$ and $\sigma_2/\sigma_1$.

The angle variable $\alpha_j$ was originally 
introduced to study the thermal fluctuations 
of the hexagonal lattice structure  around the  melting 
transition in 2D \cite{NelsonTEXT}. 
  We have  recently studied 2D  melting 
 numerically with the aid of 
visualization of $\alpha_j$ and $D_j$  
\cite{melt}, where polycrystalline  
patterns  are apparent at  the melting  not in  accord with  
the original theory \cite{NelsonTEXT} (see the end of the last section 
for more discussions).

\subsection{Bond breakage}

The jamming dynamics can 
be conveniently visualized 
if use is made of the  bond breakage \cite{yo,yo1}. 
For each particle 
 configuration  at a time $t$, 
a pair of particles $i\in\alpha$ and 
$j\in\beta$ is considered to be bonded if
\begin{equation}
r_{ij}(t) = |\bm{r}_i(t)-\bm{r}_j (t)| \le A_1
\sigma_{\alpha\beta},
\end{equation}
where $\sigma_{\alpha\beta}= (\sigma_{\alpha}+ 
\sigma_{\beta})/2$.  We set $A_1=1.2$; then, 
$A_1\sigma_{\alpha\beta}$ 
is slightly larger than the peak distance of the pair correlation 
functions $g_{\alpha\beta}(r)$. 
 After a time interval $\Delta  t$, 
the bond is regarded
to be broken if 
\begin{equation}
r_{ij}(t + \Delta t)\ge A_2\sigma_{\alpha\beta},
\label{eq:BBOD2}
\end{equation}
where we set 
$A_2=1.5$.   
In the following figures, 
bonds broken 
during a time interval $[t_1,t_2]$ 
will be marked by $\times$ 
at the middle point 
${\bi R}_{ij}= 
\frac{1}{2}({\bi r}_i(t_2)+{\bi r}_j(t_2))$ 
of the two particle positions 
at the terminal time $t_2$.  The structure factor of 
these middle points is written as   $S_b(k)$  in Eq.(1).

 In the original 
simulation using the soft-core potential\cite{yo,yo1}, 
 the bond breakage rate  was    
$
\tau_b(T,\gdot)^{-1} \cong 0.1 \tau_\alpha(T,\gdot)^{-1}   
$, 
where  $\tau_\alpha(T,\gdot)$ is 
the decay time of 
the self-correlation function and behaves as in Eq.(2) 
in shear. The 
$\tau_b(T,\gdot) $ represents  the average bond life time 
and  the broken bond number 
$\Delta N_b$ of  the whole 
system  is of order  $N\tau_b(T,\gdot)^{-1}\Delta t$ 
in  time interval $\Delta t$.  
For Wi$>1$,  
the bond breakage is mostly induced by  shear 
and the   time  average of the broken bond 
number  per particle $\Delta N_b/N $   is   
of the order of the average  strain 
 $\gdot \Delta t$ in the plastic flow regime.  
Around this mean value,  $\Delta N_b/N $  
 increases significantly   on occurrences of   
large-scale plastic events  and decreases in ``elastic periods'' 
without them \cite{preprint1}.

\subsection{Averaged velocity}

Shear bands have been  observed widely in complex fluids, 
glassy fluids, and granular materials.  
To examine  such 
strain localization in shear flow, we 
 may introduce an  averaged velocity 
 $\bar{v}_x(y,t)$ as follows \cite{preprint1}. 
We first integrate  
the $x$ component of the momentum density  
${J}_x(x,y,t)$ 
 in the flow  direction:  
\be
{\bar J}_x(y,t)= \frac{1}{L} 
\int_{-{L}/{2}}^{{L}/{2}} 
 dx J_x(x,y,t)
=  \frac{1}{L}  \sum_j m_j {\dot x}_j(t) 
 \delta (y-y_j(t)).
\en  
We  next  smooth ${\bar J}_x(y,t)$ 
over  space and time intervals with widths $\Delta y$ 
and $\Delta t$ as  
\begin{equation}
{\bar v}_x(y, t) = 
\frac{1}{{\bar \rho}\Delta y\Delta t}
\int^{y+{\Delta y}/{2}}_{{y}-{\Delta y}/{2}}
\hspace{-1mm} {dy'} 
\int^{t+\Delta t}_{t} \hspace{-2mm} {dt'} 
\bar{J}_x(y',t'),
\label{eq:veloc_prof}
\en 
where ${\bar \rho}$ 
is the average mass density. We fix $\Delta y$ at $L/20$. 
However,   
we choose $\Delta t$ depending on the time scale 
of plastic events under consideration. 
This is because of   the hierarchical dynamics of   plastic deformations.

\section{Numerical results in quiescent states}

Without applied shear,  
we first integrated  
the equations of motion under 
the periodic boundary conditions in the $x$ and $y$ directions. 
We  equilibrated the  system at 
$T=0.2$  for a time interval of $5\times 10^3$ until  
 no appreciable time evolution was  detected in 
various thermodynamic quantities. At this  time we will 
 set $t=0$. We attached a Nos\`e-Hoover thermostat \cite{nose,Hoover} 
to all the particles, which then obeyed   
\be 
m_j \ddot{\bm{r}}_j =  
- \frac{\partial }{\partial\bi{r}_j} U - 
\zeta  m_j \dot{\bi r}_j, 
\en 
where   $\dot{\bi r}_j= d{\bi r}_j/dt$,   
$\ddot{\bi r}_j= d^2{\bi r}_j/dt^2$, and 
 $U$ is the total potential. The 
 thermostat variable  $\zeta(t) $  obeyed      
\be
\frac{d}{dt} {\zeta} =  \frac{1}{\tau_{\rm{NH}}^2}
\bigg[ \frac{1}{N T}\sum_{j } 
\frac{m_j}{2} {\dot{\bi  r}_j^2} -1\bigg], 
\en
where  $\tau_{\rm{NH}}$ is the thermostat 
characteristic time. 
We set $\tau_{\rm{NH}}= 0.961$. 

\begin{figure}[b]
\centerline{\includegraphics[height=7.4 cm]{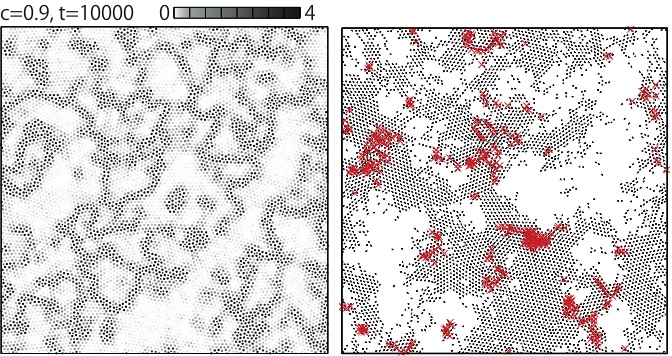}}
  \caption{ (color online) Left: 
disorder variables $D_j (t)$ at $t=10^4$ 
with the gradation bar at the top, where $\gdot=0$  and  $c=0.9$. 
Crystalline regions composed of the large particles (white) 
are enclosed by  percolated amorphous layers containing the small 
particles (gray or black). 
Right: broken bonds ($\times$) and 
displacement vectors  for mobile 
particles with  $|\Delta {\bi r}_j|>0.2$  
 in the  time interval $[10^4,4\times 10^4]$ with width 
$\Delta t=3\times 10^4$ in the same run. 
The particles with $D_j>4$ are 
written in black (left).  The whole  system $-L/2<x,y<L/2$ 
is shown. 
}
\end{figure}
\begin{figure}
\centerline{\includegraphics[height= 9.4 cm]{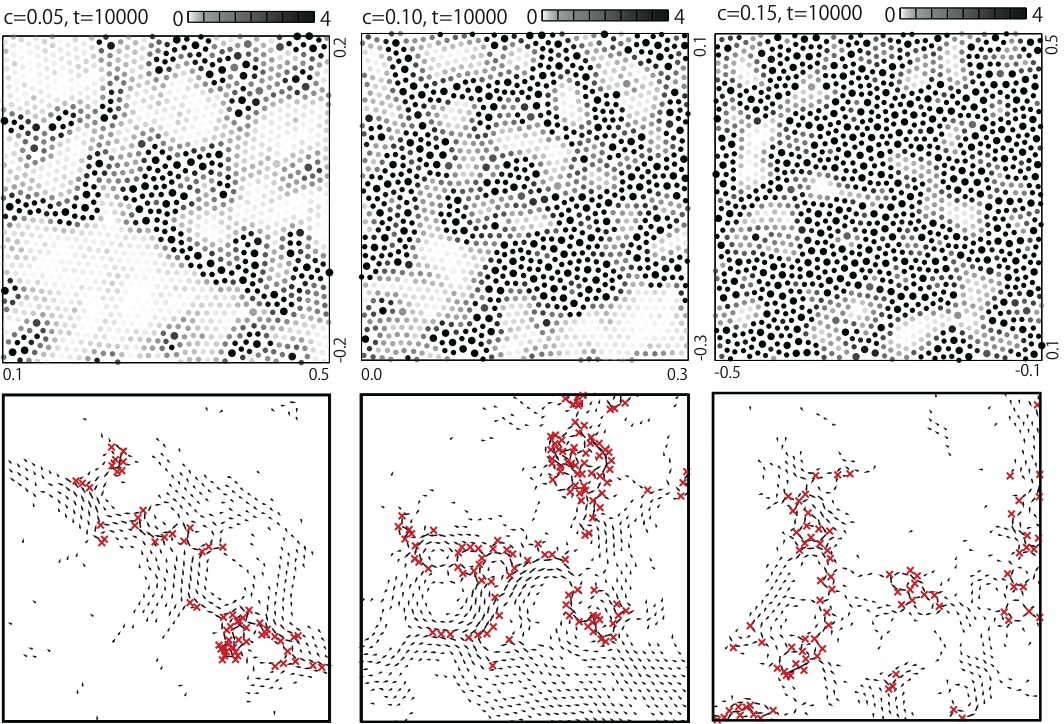}}
   \caption{ (color online) Upper plates: disorder variables $D_j(t) $ at $t=10^4$ 
 for  $c=0.05$, $0.1$, and $0.15$ with $\gdot=0$. 
Lower plates: broken bonds ($\times$) and 
displacement vectors  for mobile particles with  $|\Delta {\bi r}_j|>0.2$    
 in the  time interval $[10^4,4\times 10^4]$ 
 in the same run producing the corresponding upper  
panel. These are snapshots of 
$1/16$ of the total system.  The numbers at the corners 
(upper plates) denote the normalized coordinate $x/L$ or  $y/L$.  
}
   \label{fig:3}
   \end{figure}

\begin{figure}
\centerline{\includegraphics[height=9.4 cm]{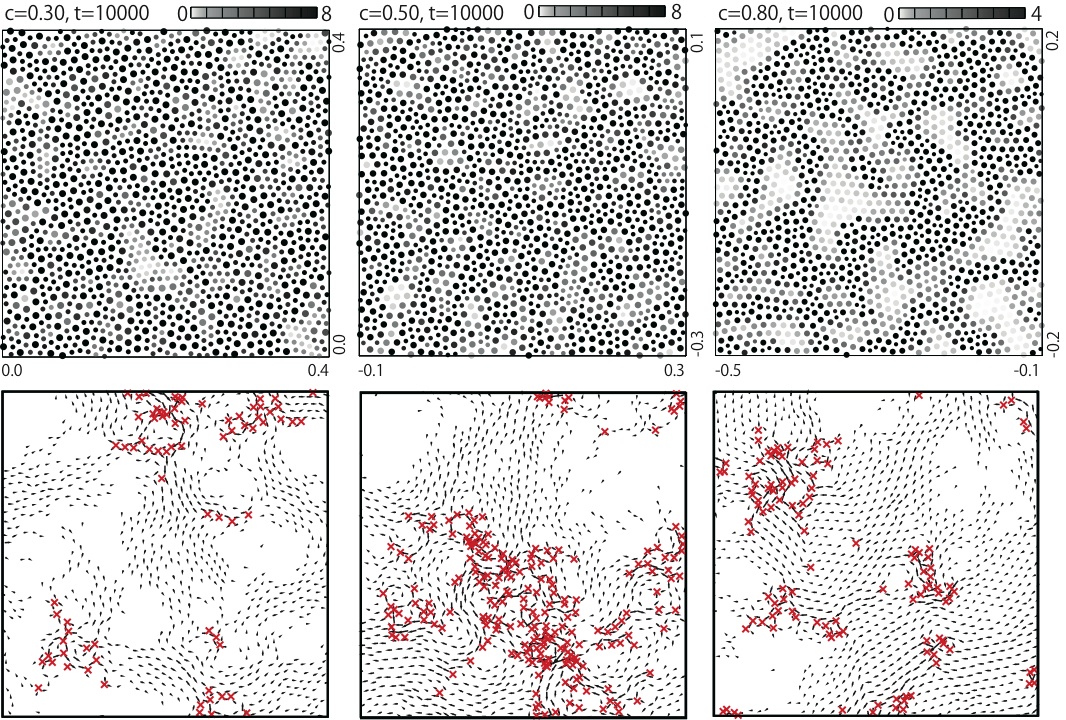}}
   \caption{ (color online) Disorder variables $D_j(t)$ (upper plates),  
and broken bonds ($\times$) and 
 displacement vectors  for mobile  particles with  $|\Delta {\bi r}_j|>0.2$  
(lower plates)   
for $c=0.3$, $0.5$, and $0.8$ with $\gdot=0$. 
   For $c=0.3$ and 0.5  
the  particles  with $D_j>8$  are 
written in black (upper plates).}
   \label{fig:4}
   \end{figure}
\begin{figure}
\centerline{\includegraphics[width=14 cm,height=7.5 cm]{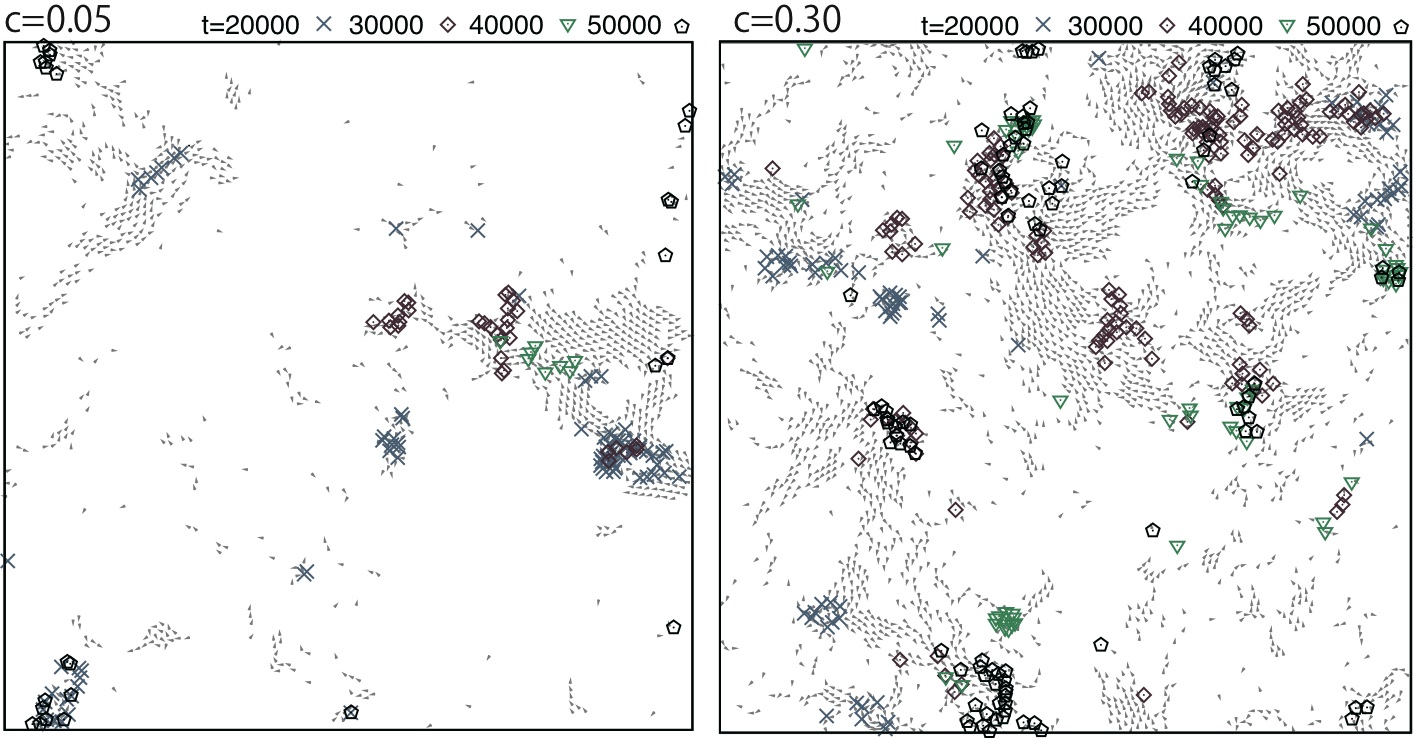}}
   \caption{ (color online) Broken bonds 
and particle displacements  with  $|\Delta {\bi r}_j|>0.2$  
(arrows) in consecutive four 
time intervals with width $\Delta t=10^4$ 
for $c=0.05$ (left) and $0.3$ (right) 
with $\gdot=0$ in the whole system. Shown is  time evolution of 
the dynamic heterogeneity in polycrystal (left) and glass (right).}
   \label{fig:5}
   \end{figure}

By varying $c$ in the range $c \le 0.5$ 
 at  $\sigma_2/\sigma_1=1.4$, 
Hamanaka and one of the present authors \cite{hama,HamaOnuki} 
 realized  crystal states 
with small numbers of defects 
for $c\ls 0.01$, polycrystal states 
with the large particles 
forming grain boundaries for $0.01\ls c\ls 0.12$, 
and glass states for larger $c$.    For  $c\gs 0.1$,  
the system is  divided into  
small crystalline  domains  composed of the small  particles  
and percolated amorphous regions composed of 
the small and  large particles. 
The areal fraction of the amorphous regions 
increases  with increasing $c$.   See Fig. 2 of 
Ref.14 for the particle configurations for various 
$c$  at  $\sigma_2/\sigma_1=1.4$, where  
 a crystal state was 
 realized for $c=0.02$ with 
large particles being distributed as point defects.

In this work, we carried out many  simulation runs also  
 for $c>0.5$. For not very small  $c$ or $1-c$,   
the system is divided into  crystalline 
regions composed of the large  particles 
and amorphpus layers composed of the two species. 
The areal fraction of the layers 
is decreased for relatively small  $c$ or $1-c$.  
  Figure 2 provides such an example  for  $c=0.9$,   
where we  display 
 $D_j(t)$ at $t=10^4$ in the left and  
the broken bonds and the displacement vectors 
\be 
\Delta {\bi r}_j(t,\Delta t)=
{\bi r}_j(t)-{\bi r}_j(t-\Delta t),   
\en  
in the  time interval $[10^4,4\times 10^4]$ with width 
$\Delta t= 3\times 10^4$ in the right.  In writing the displacements 
 we   pick up only mobile 
particles  with  $|\Delta {\bi r}_j|>0.2$. 
At this composition   
percolation of the  amorphous layers 
is attained.  However, for $c=0.95$, 
 the layers  are broken into 
isolated clusters with various sizes, 
resulting in a crystal state with point defects  (not shown here).  
 In Fig. 2,  the broken bonds are mostly 
created  by the particle motions in 
the  layers and are  
heterogeneously distributed. 
We also find coherent  motions   of  crystalline 
domains  surrounded by    broken bonds    
at their  grain boundaries. See  
the middle part at the bottom in the two panels 
for such an  example. The patterns 
should be  isotropic under the periodic boundary condition 
(on the average of many runs). 
Next, in Figs. 3 and 4, we show the same quantities 
for $c=0.05$, $0.1$, $0.15$, $0.3$, $0.5$, and $0.8$ 
to illustrate the crossover with increasing $c$. 
The system is in a well-defined  polycrystal state for $c=0.05$ 
and is  highly disordered for $c=0.3$ and 0.5. 
The structural heterogeneity in $D_j$ 
and the dynamical one in the broken bonds and 
the displacements  are both conspicuous   
for any $c$. Their correlation is obviously seen  
in the presence of small crystalline regions. 
This was   illustrated in Ref. 14 in  more expanded snapshots 
 including the   elusive  case of glass.

Furthermore, in Fig. 5, we show  time evolution  of the broken 
bond distributions in four consecutive  time intervals with width $10^4$ 
in polycrystal at $c=0.05$ and in glass at $c=0.3$. 
The structural relaxation time $\tau_\alpha$ is of order 
$10^5$ at $c=0.05$ and  $10^4$ at $c=0.3$\cite{hama}. 
Here  the clusters of the broken bonds mostly 
overlap or are adjacent to each other, resulting in aggregations 
of the broken bonds. Similar figures illustrating the heterogeneity 
evolution can be found in other papers also 
\cite{yo,yo1,Dol,preprint2}.

\section{Numerical results at $\gdot=10^{-4}$}

We  applied  a simple shear flow with shear rate $\gdot=10^{-4}$.  
To this end, we divided  the system into three 
regions \cite{HamanakaShear,preprint1}. 
 In  the bulk region  $-0.5L<x,y<0.5L$, 
we initially placed 
particles with $N=9000$.  
We  added    two boundary layers 
 in the region   $-0.6L<y<-0.5 L$ at the bottom   
and  in the region  $0.5L<y<0.6 L$ at the top.  
In  each layer, 
$N_b= 900$  particles
with the same composition and size ratio 
were initially placed. They  were  attached to the layer  
by the spring potential,  
\begin{equation}
u_{j} (\bi{r}-\bi{R}_j) 
= \frac{1}{2} K |\bi{r} -\bi{R}_j|^2, 
\label{eq:SPP1}
\end{equation} 
where   $\bi{R}_j=\bm{R}_j(t)$ are 
pinning points  in the  boundary layers dependent on $t$. 
The spring constant was  set equal to $K= {20\epsilon}{\sigma_1^{-2}} $.
These bound particles  also interacted  with the neighboring  
bound and unbound particles with the common 
Lennard-Jones potentials in Eq. (\ref{eq:LJP}). 
The $x$ component 
of $\bm{R}_j$  was moved  as 
\begin{equation}
X_j (t)= X_j(0)\pm \frac{1}{2}L\dot{\gamma}t .
\label{eq:FPP1}
\end{equation}
We imposed  the periodic boundary condition in the $x$ direction. 
When   $X_j(t)>L/2$ (or  $X_j(t)<-L/2$) in the integration, 
$X_j(t)$  was decreased (or increased) by $L$. 
The unbound particles rarely penetrated   into 
the boundary layers deeper than  $\sigma_1$.

The unbound particles in the bulk regions obeyed  the 
Newtonian equations of motion without thermostat. 
However, a Nos\`e-Hoover thermostat \cite{nose,Hoover} was 
attached 
\begin{wrapfigure}{l}{6.6cm}
\centerline{\includegraphics[width=7.1 cm,height=5 cm]{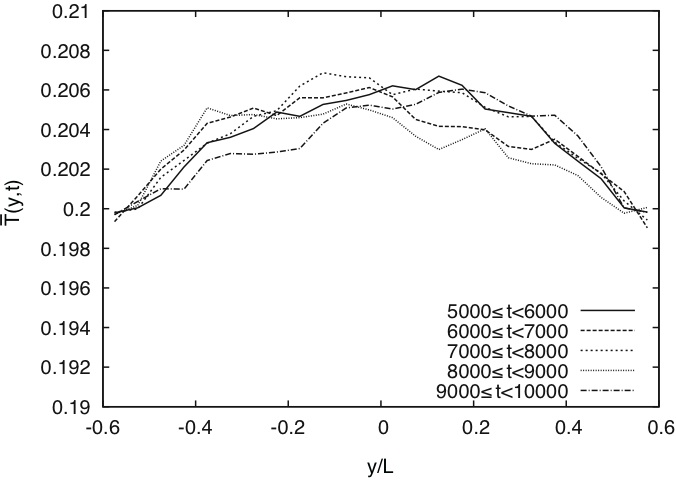}}
 \caption{Averaged temperature  $\bar{T}(y,t)$ vs $y/L$ 
with  $\gdot=10^{-4}$  for $c=0.05$, which is higher 
in the middle than the boundary value 
$0.2$ only by  a few $\%$. 
 }
\end{wrapfigure}  
 to each boundary layer independently. That is, 
 the bound particles $j\in {\cal B}$
were governed by   
\be 
m_j {\ddot {\bm{r}}}_j =  
- \frac{\partial }{\partial\bi{r}_j}U  - 
\zeta_{\cal B}   m_j (\dot{\bi r}_j-{\bi v}_{\cal B}),
\en 
where $\cal B$ denotes the top or the bottom. 
The boundary velocity ${\bi v}_{\cal B}$ 
is $(L\gdot/2){\bi e}_x$ at the top 
and  $ -(L\gdot/2){\bi e}_x$ at the bottom, where 
 ${\bi e}_x$ is  the unit vector along the $x$ axis.  
 For $\gdot=10^{-4}$ 
the   boundary speed ($\sim 0.005$) 
is much slower than the thermal velocity 
($\sim 0.6$).   The two thermostat parameters,  $\zeta_{\rm bot}$ 
at  the bottom and $\zeta_{\rm top}$ 
at  the top, obeyed       
\be
\frac{d}{dt} {\zeta}_{\cal B} =  \frac{1}{\tau_{\rm{NH}}^2}
\bigg[ \frac{1}{N_b T}\sum_{j \in {\cal B} } 
\frac{m_j}{2} {|\dot{\bi  r}_j-{\bi v}_{\cal B}|^2} -1\bigg]. 
\en
We set $\tau_{\rm{NH}}= 0.304 $ and $T=0.2$. 
For $\gdot=10^{-4}$,  the local temperature  
was  nearly homogeneous  in the bulk region 
during plastic flow, as demonstrated  in Fig. 6 for $c=0.05$. 
Setting  $\Delta t=10^3$ and $\Delta y=L/20$,  
we define  the averaged temperature  by 
\be 
{\bar T}(y,t)= \frac{1}{L \Delta y\Delta t}\int_{t-\Delta t}^t  dt' 
\int^{y+{\Delta y}/{2}}_{{y}-{\Delta y}/{2}}dy'\int_{-L/2}^{L/2}dx' 
\epsilon_K (x',y',t'), 
\en 
where $\epsilon_K (x,y,t)=\sum_j m_j |{\dot{\bi R}}_j-
\gdot y {\bi e}_x|^2\delta ({\bi r}- {\bi R}_j)/2 $ 
is the kinetic energy density.

\begin{figure}
\centerline{\includegraphics[height=12 cm]{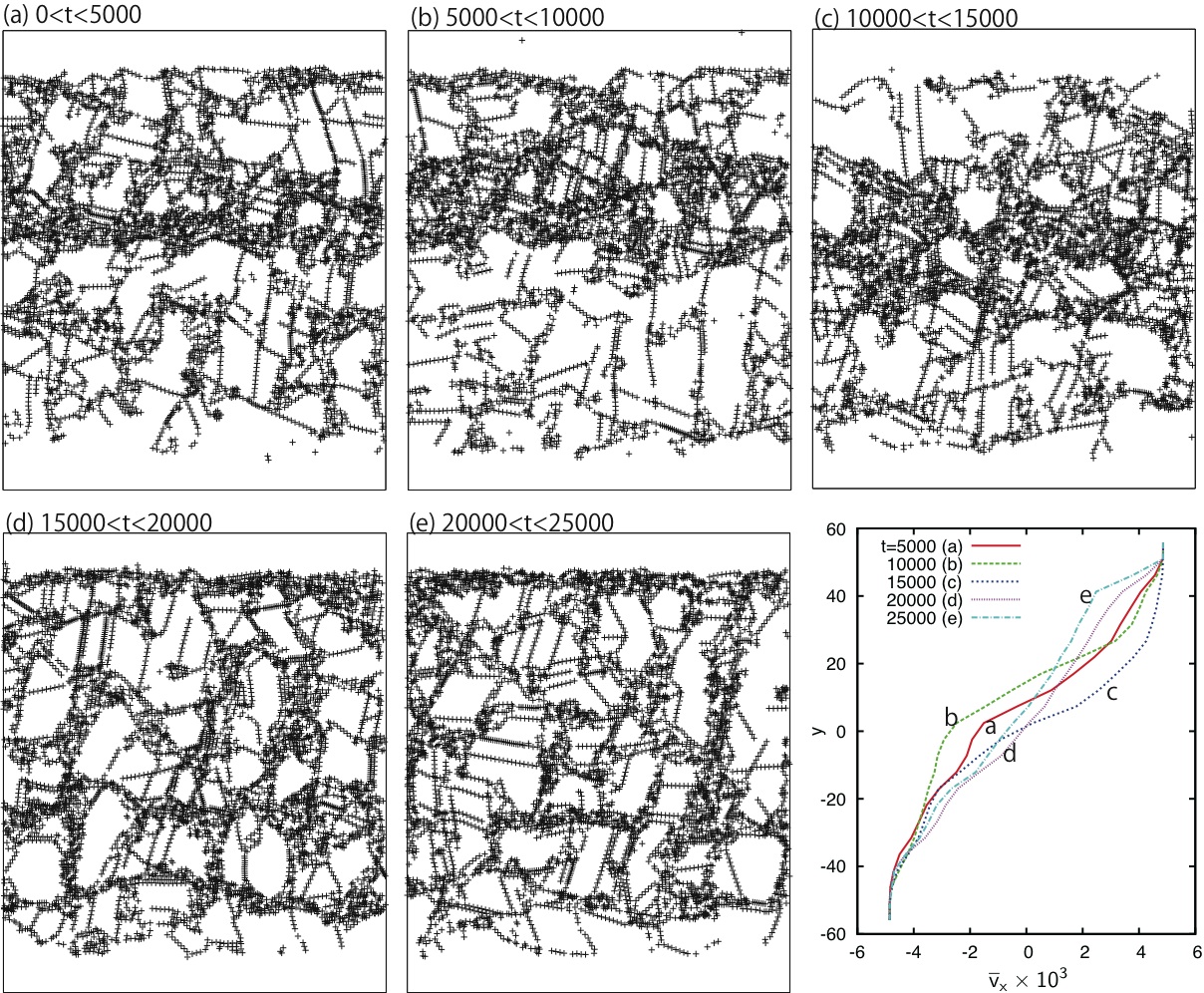}}
   \caption{Broken bonds in five  consecutive time intervals 
with width $\Delta t=5000=0.5/\gdot$ 
in shear $\gdot=10^{-4}$ in polycrystal with 
 $c=0.05$. They form slip lines connecting grains. 
The top and bottom boundary layers 
are without broken bonds. 
Right bottom: averaged velocity ${\bar v}_x(y,t)$ defined by 
Eqs. (12) and (13) from the same data for these 
time intervals, which greatly 
changes in time and much deviates from the linear profile.}
   \label{fig:6}
   \end{figure}

We  prepared 
 the initial states  as follows. 
First, we  equilibrated the bulk 
and boundary  regions independently without their mutual interactions. 
The particles in these  regions interacted 
via the LJ  potentials  in Eq.(\ref{eq:LJP})  
 in a liquid state at $T=2$ in 
a time interval of $10^3$ under  the periodic 
boundary condition along   the $x$ and $y$ axes. 
Second,  we quenched the  system to 
$T=0.2$ and further equilibrated it  
 for a time interval of $10^3$. 
Afterwards, we chose the  particle  positions in the 
boundary layers  as  the initial pinning   points 
${\bi R}_j(0)$  and 
 introduced the spring potential  
of the bound particles and 
the LJ potentials  between 
 the bound and unbound particles.  
Third, we  further waited  for time interval of 
$5\times 10^3$ until we detected 
 no appreciable time evolution in 
various  quantities. After this second low-temperature equilibration, we 
 applied a  shear flow with  rate 
$\dot{\gamma} = 10^{-4}$ by
sliding the pinning positions 
in the  boundary layers   
with velocities  $\pm \dot{\gamma}L/2$.
We  set $t=0$ at the application of shear.

\begin{figure}
\centerline{\includegraphics[height=12 cm]{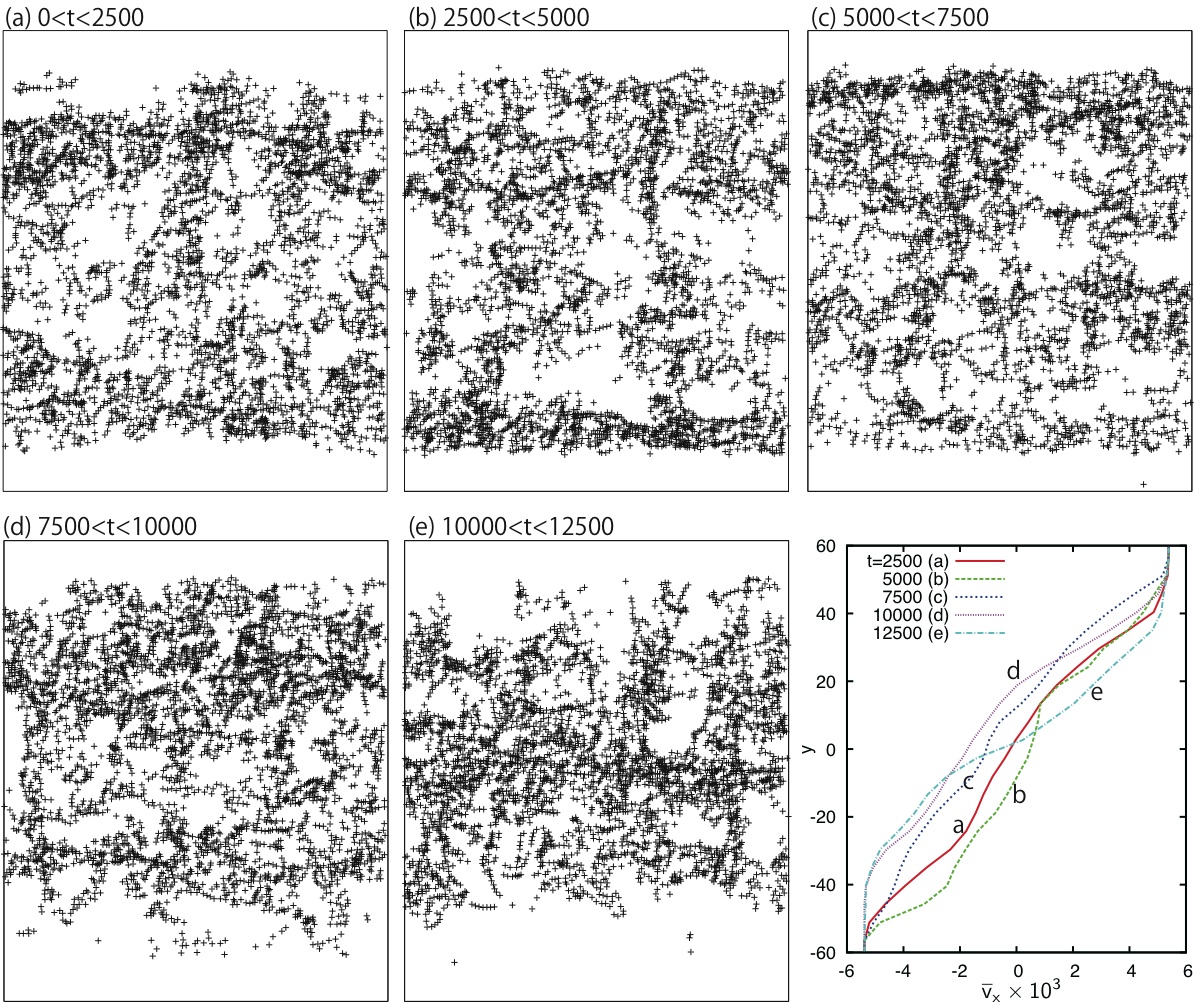}}
   \caption{Broken bonds in five  
 consecutive time intervals 
with width $\Delta t=2500=0.25/\gdot$ in shear $\gdot=10^{-4}$ 
in glass with  $c=0.3$. Slip lines are short. 
 Right bottom: averaged velocity ${\bar v}_x(y,t)$ 
from the same data for these 
time intervals.}
   \label{fig:7}
   \end{figure}
\begin{figure}
\centerline{\includegraphics[height=13.8 cm]{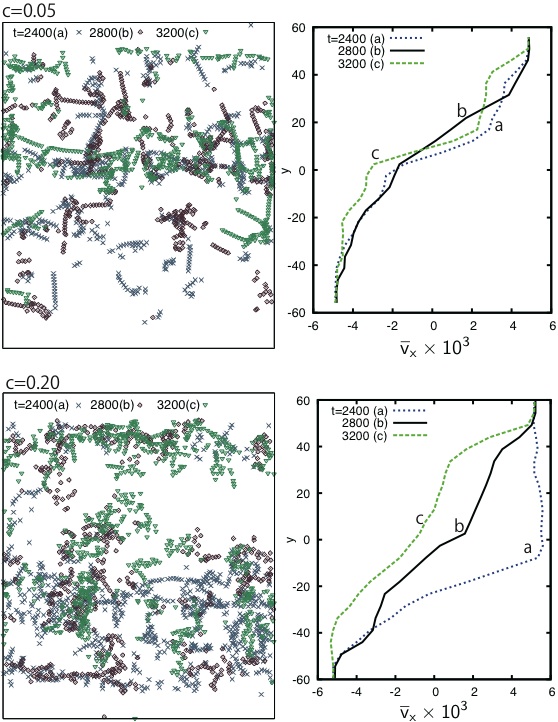}}
\caption{ (color online) Left: time evolution of bond breakage 
in three consecutive time intervals 
with width $\Delta t= 400=0.04/\gdot$ in shear $\gdot=10^{-4}$. 
 Right: averaged velocity ${\bar v}_x(y,t)$ for these 
time intervals much deviating from   the linear profile. 
The system is  in polycrystal  with  $c=0.05$ (top) and 
in glass with $c=0.2$ (bottom). Repeated occurrences of 
plastic events result in transient shear bands.}
   \label{fig:8}
   \end{figure}

In Fig. 7,  we show the broken bonds 
in five  consecutive time intervals 
with width $5000=0.5/\gdot$ in  polycrystal at $c=0.05$.  
Lines of the  broken bonds represent plastic events 
created by dislocation gliding \cite{Friedel} 
or grain boundary sliding \cite{Yip}. 
The former deformation  mode is dominant at this small $c$, 
while the latter  becomes increasing important 
with increasing $c$ \cite{preprint1,Yip}.  We can see that 
most of the slip lines   are nearly  
 parallel to  the $x$ or $y$ axis.  This is 
because  the elastic energy of a slip 
is minimum in these directions in simple shear strain 
\cite{Onuki-plastic}.  
In the first three time intervals 
 a thick  band region extends over the system 
along the $x$ axis 
in the upper part, but in the last two 
time intervals such large-scale bands are nonexistent. 
The  right bottom panel 
displays the average velocity ${\bar v}_x(y,t)$ 
 defined in Eqs. (12) and (13) with $\Delta t= 5\times 10^3$, 
where the first three curves largely deviate from the linear 
profile $\gdot y$ but the last two curves are nearly linear.

In Fig. 8, we show the same quantities 
in five consecutive time intervals 
with width $2500=0.25/\gdot$  in glass  at  $c=0.3$. 
Also in this    case,  the plastic deformations 
are highly heterogeneous, 
which still  tend to be   nearly parallel to  the $x$ or $y$ axis.  
The chains of the broken bonds are shorter 
than in Fig. 7. That is, the slip lines do not much 
exceed the molecular size and 
the broken bonds are more randomly scattered than in polycrystal. 
 As a result, the deviation 
of the averaged velocity ${\bar v}_x(y,t)$  
 from the  linear form 
becomes small for $\Delta t\gs 1/\gdot$, 
while it is large for $\Delta t\ls 0.1/\gdot$ (see Fig. 9). 
In their long time simulation 
of sheared 2D glass, 
 Furukawa {\it et al.}\cite{Furukawa} 
 realized  transient shear  bands  
equally in the flow and velocity-gradient 
directions  with  the Lees-Edwards  boundary condition,   
while   those parallel to the $x$ axis  more 
easily  extend  upto the system length  
in our simulation with the boundary layers.

 Figure 9 illustrates   the time-evolution 
of the dynamic heterogeneity  
  in three consecutive  time intervals 
with width $400=0.04/\gdot$  
in polycrystal at $c=0.05$ and in glass at $c=0.3$. 
We display  the broken bonds  
in the  left panels 
and the averaged velocity 
${\bar v}_x(y,t)$  in the right  panels. 
The width  $\Delta t$ here is made much shorter than 
in Figs. 7 and 8 to analyze the  early stage of plastic deformations. 
Note that the transverse sound velocity $c_\perp$ 
is of order 5 (in units of $\sigma/\tau$)  
and the acoustic traversal time $L/c_\perp$ 
is of order 20  in our system. 
(Plastic deformation dynamics on shoter 
time scales has been analyzed in our previous work \cite{preprint1}.) 
 In the upper panels at $c=0.05$,   
large-scale plastic deformations 
multiply  occur  in the same area, which extends over the system 
to   form  a shear band. On the other hand, 
in the lower  panels at $c=0.3$, 
a large shear band is seen 
in the lower part in the first time interval,  but  
  large plastic 
deformations occur in the vertical direction in the second time interval 
and  in the upper part in the horizontal direction in the third time interval. 
The life time of the 
shear bands is  shorter in glass than in polycrystal \cite{preprint1}.
In addition, the  broken bond numbers 
$\Delta N_b$ in the whole  system  
are 649,    551, and 614 for $c=0.05$ 
and  are 870,   589, and  761 for $c=0.3$ 
in these  time intervals  in the chronological  order. 
The time average of $\Delta N_b$ is 
 of order $N\gdot\Delta t (=360$ here), 
which should be  the case  
for any $c$ as discussed at the end of Subsec.2.3 \cite{preprint1}.

\section{Summary and remarks}

We have examined the   jammed particle configurations 
and the dynamic  heterogeneity in 2D.  
Visualization of 
the disorder variable $D_j$ gives 
information of the structural heterogeneity, 
while that of  the bond breakage and 
the particle displacements 
discloses  the presence of the  dynamic heterogeneity. 
We have 
varied  the composition $c$ for 
$\sigma_2/\sigma_1=1.4$ and $T=0.2\epsilon/k_B$. 
Summarizing  our main results, we give some remarks below.

The particle configurations are unique  
in polycrystal and glass which are realized 
for not very small $c$ or $1-c$ in our 2D model system.  
As in Figs. 2-4, small crystalline regions composed of 
one species of the particles are enclosed 
by  percolated  amorphous regions 
composed of the two species, where the amorphous 
 regions  form layers for 
 relatively small $c$ or $1-c$. For 
very small  $c$ or $1-c$ (not shown in this work), 
the layers  break  into small 
pieces to form a crystal with point defects. 
In rheology, these  amorphous layers 
can serve as a lubricant in plastic flow  
\cite{preprint1}, within which 
the bond  breakage occurs, 
reducing the viscosity than 
in crystal states.

In glass without  shear,  the  
 particle configuration 
changes are thermally induced  in the form of  
 chains of broken bonds \cite{yo,yo1} 
or stringlike displacements \cite{Kob}. 
They   accumulate to form the dynamic heterogeneity 
on long times  characterized by 
the correlation length $\xi$ in Eq.(1) 
\cite{yo,preprint1,preprint2}. 
 In Figs. 2-4, we have  added more   
 evidences for  these dynamical processes. 
To show the  relationship between 
 the structural and dynamical heterogeneities,  
we have presented the snapshots of the disorder variable 
$D_j$ and those of the bond breakage and the 
displacements (see  Fig. 2 
in Ref.14 also). 
In Fig. 5, we have visualized 
the time evolution of the dynamic heterogeneity 
on the time scale of $10^4$, which demonstrates  
correlated occurrences of mesoscopic  
structural   changes. This tendency  is    
 related  to the mesoscopic heterogeneity 
in the  elastic moduli \cite{Yoshimoto,Barrat-small}. 
In  addition, in Fig. 4, the structural relaxation 
is even slower   in polycrystal than in glass, where 
 the particles around the grain boundaries 
(in the amorphous layers) are relatively mobile 
than those within the crystalline grains\cite{HamaOnuki}.

In shear flow with $\gdot=10^{-4}$,  
we have visualized large-scale 
heterogeneity in   the bond breakage in Figs. 6-8. 
The  averaged velocity 
${\bar v}_x(y,t)$  deviates from the linear 
profile with formation of shear bands. 
 Here    $\Delta t= 0.5/\gdot$  for   $c=0.05$ in Fig. 7, 
 $\Delta t= 0.25/\gdot$  for  $c=0.3$ in Fig. 8, 
 and $\Delta t= 0.04/\gdot$  for  $c=0.05$ and $0.3$ in Fig. 9. 
These figures illustrate how the plastic deformations under shear 
evolve  on various  time scales.  
For any $c$, plastic deformations 
extend over longer distances 
nearly in the $x$ or $y$  axis  as in Fig. 9. 
Such plastic events often accumulate to form transient 
shear bands as in Figs. 7-8. 
They occur to release the elastic energy 
at high strain as discussed in Sec.I. 
See Refs. 27 and 28  for more  discussions 
on the anisotropic, hierarchical 
 dynamics under  shear. 
In the early  work  by Yamamoto and one of the present authors\cite{yo1},   
the shear-induced structural change was analyzed on the basis of 
 Eq.(2) and the broken-bond structure factor 
$S_b(k)$ in Eq.(1) was calculated 
for $\Delta t=0.05\tau_b\cong 
0.5\tau_\alpha$  without and with  shear,  
but the shear band formation on large scales 
was beyond its scope.

A variety of complex 
problems of binary particle systems  remain mostly unexplored. 
We mention some of them.  
(i) When  the size ratio $\sigma_2/\sigma_1$ 
is increased from unity, intriguing  crossovers are expected   
in  the particle configurations 
and the dynamics. For example, at $c=0.5$, 
  proliferation of defects occurs abruptly 
 around  $\sigma_2/\sigma_1\cong 1.2$ \cite{hama,HamaOnuki}. 
(ii) Furthermore, with changing the pair potentials,  
tendency to phase sparation can be enhanced, where 
small crystalline regions should become 
more disinct or  nucleation of crystal 
domains should become realizable in an  amorphous matrix.   
(iii) Some essential aspects of the glass problem 
should be common in 2D and 3D.  
However, in 3D, the problem is much more complicated 
 and visualization of configuration 
changes is more difficult. 
We have not yet understood  
the differences in 2D and 3D in depth. 
(iv) In addition, the melting and  crystallization 
in binary mixtures have  not yet been well investigated, 
where essentially 
different pictures arise in 2D and 3D.  
We have recently recognized 
surprising complexity of 
the 2D melting even  in one component systems\cite{melt}.  
That is, marked   heterogeneities 
emerge both  in the structural disorder 
and in the dynamics  
 in the  hexatic phase  \cite{NelsonTEXT}, 
where mesoscopic liquidlike and crystalline  regions coexist 
as thermal fluctuations without distinct interfaces. 
 Understanding the dynamics of 2D melting is still at the beginning  
 despite 
numerous papers on static  properties.

\section*{Acknowledgements}

The authors would like to thank Akira Furukawa and 
Ryoichi Yamamoto 
for valuable discussions. 
Some of the numerical calculations were carried out
on Altix 3700B at ISSP Supercomputer Center, Univ. of Tokyo. 
This work was supported by Grants-in-Aid 
for scientific research 
on Priority Area ``Soft Matter Physics" 
and  the Global COE program 
``The Next Generation of Physics, Spun from Universality and Emergence" 
of Kyoto University 
 from the Ministry of Education, 
Culture, Sports, Science and Technology of Japan. 

%

\end{document}